# TiFe$_{0.85}$Mn$_{0.05}$ alloy produced at industrial level for a hydrogen storage plant

Jussara Barale,[a] Erika M. Dematteis,[a)b)] Giovanni Capurso,[c)1] Bettina Neuman,[d] Stefano Deledda,[e] Paola Rizzi,[a) *] Fermin Cuevas,[b] and Marcello Baricco[a]

a) Department of Chemistry and NIS - INSTM, University of Turin, Via Pietro Giuria 7, 10125 Torino, Italy.
b) Univ. Paris Est Creteil, CNRS, ICMPE, UMR7182, 2-8 rue Henri Dunant, 94320 Thiais, France.
c) Institute of Hydrogen Technology, Helmholtz-Zentrum Hereon, Max-Plank-Straße 1, 21502 Geesthacht, Germany.
d) GKN Sinter Metals Engineering GmbH, Krebsöge 10, D-42477 Radevormwald, Germany
e) Institute for Energy Technology, Instituttveien 18, 2007 Kjeller, Norway.
[1] Present address: Department of Polytechnic Engineering and Architecture, University of Udine, Via Cotonificio 108, 33100 – Udine, Italy

*Corresponding author
Paola Rizzi
E-mail address: paola.rizzi@unito.it

Supplementary Tables and Figures available.

**Abstract**

In the use of hydrogen as large-scale energy vector, metal hydrides based on intermetallic compounds play a key role, thanks to mild temperatures and pressures required for the storage. It is increasingly necessary to develop hydrogen-based devices, and in this work, the intermetallic compound TiFe$_{0.85}$Mn$_{0.05}$ is evaluated and selected as H$_2$-carrier for a storage system of 50 kg of H$_2$. A batch of 5 kg of alloy was synthesized at industrial level and characterized through scanning electron microscopy and powder X-ray diffraction, to determine the structure and phase abundance. Moreover, the hydrogen sorption properties were investigated through thermodynamic and kinetic analyses, followed by a long-term cycling study and resistance to O$_2$ and H$_2$O poisoning. Comparing results for alloys with same nominal composition, but prepared either under industrial or laboratory conditions, it was found that the alloy synthesis promotes discrepancies in phase abundance and microstructure and promotes the formation of a passive layer that deeply affect the hydrogen sorption properties. A scheme based on Monte Carlo simulation and structural results was developed to explain the key role of the passive layer (Ti$_3$Fe$_3$O) and of the secondary phases (Ti$_4$Fe$_2$O$_{0.4}$ and β-Ti$_{80}$(Fe,Mn)$_{20}$) in promoting the hydrogenation of the TiFe$_{0.85}$Mn$_{0.05}$. A storage system based on this alloy can be integrated with an electrolyser upstream (25 bar) and a fuel cell downstream (1 bar) at 55 °C, storing 1.0 H$_2$ wt.%, displaying fast kinetic, resistance to oxygen, water and nitrogen gas impurities, and stability over more than 250 cycles.

## 1. Introduction

Hydrogen is a promising widespread energy vector, and its efficient storage is a crucial step for its usage in large scale. Metal hydrides (MHs) allow storing H$_2$ at mild conditions, *e.g.* ≤ 30 bar and room temperature, resulting in a more efficient and safer solution than compressed and liquefied H$_2$. A huge number of alloys has been investigated as hydrides forming materials[1–3], but only few storage devices are commercially available nowadays. Bellosta von Colbe et al. reported MH-storage systems for stationary and mobile applications, presenting future perspectives and highlighting their potential in stationary uses for smart-grid and off-grid energy management[4]. Intermetallic compounds (IMCs) are good candidates for these types of applications, since they can absorb and desorb H$_2$ at mild pressures and temperatures, being easily integrated with Proton Exchange Membrane (PEM) Electrolysers (EL) and PEM- Fuel Cells (FC). These compounds generally have low gravimetric hydrogen capacity (< 2 H$_2$ wt.%), implying a large mass of carrier needed to store kilograms of H$_2$, coupled with high volumetric density (> 80 kgH$_2$/m$^3$). Because in stationary uses there are few limitations on the system weight, but a strong request for low volumes, IMCs are suitable materials to be employed in fixed hydrogen storage systems, *e.g.* for the management of intermittent renewable energies.

The TiFe IMC has a big potential for widespread use in stationary applications, since it stores hydrogen at mild pressures and temperatures (*e.g.* 10-20 bar and 30-70 °C). Moreover, it is inexpensive, non-toxic and the raw materials are abundant[5,6]. TiFe has the cubic CsCl-type structure, space group $Pm\bar{3}m$, with a cell parameter from 2.974 to 2.978 Å, depending on Ti at.%, *i.e.* from 49.5 % to 52.5 % [6–8]. After dissolving hydrogen in solid solution (α-TiFeH$_{0.06}$), it forms two subsequent hydrides, visible in the pressure-composition-Temperature (pcT) curves through two consequent plateaux: a monohydride (β-TiFeH) and a dihydride (γ-TiFeH$_2$), presenting a maximum gravimetric capacity close to 2 H$_2$ wt.%[9]. A drawback is its hard activation, which usually requires high temperatures (*e.g.* 400 °C)[5,9,10] and/or high hydrogen pressures (up to 100 bar)[11]. A strategy to overcome this problem and to tailor H$_2$ sorption properties involves the substitution of Ti and/or Fe with other elements [12–20], and/or the enrichment of Ti in the composition[18], to promote the formation of secondary phases that have been found to help in the activation process. A large hysteresis gap between absorption and desorption pressure is usually observed, which depends on the sample history (*e.g.* synthesis, activation)[9]. A comprehensive overview of the properties of TiFe and TiFe-based alloys was reported in two recent reviews by Sujan[6] et al. and Dematteis et al.[19]. Mn is a common substitution for Fe in TiFe, forming TiFe$_x$Mn$_{(1-x)}$ alloys[21]. As reported by Johnson et al.[21], Mn containing compounds showed a decrease in the equilibrium pressures for hydrogen sorption and an increase in cell dimension, with respect to the original TiFe phase. On the contrary, no changes in gravimetric capacity were evidenced up to x = 0.70. Mn additions result in the occurrence of sloping plateaux in the pcT-curves. These can be limited by suitable annealing, as well as to a reduction of the hysteresis gap[21]. Challet et al.[22], prepared and studied Ti(Fe$_{0.70+x}$Mn$_{0.20-x}$) alloys, with x = 0.00, 0.05, 0.10 and 0.15. Samples were synthesised by induction melting, annealed at 1000 °C for 1 week and then quenched. For the first hydrogenation, activation was not necessary for any sample and pcT-curves presented two distinct plateaux with limited slope. Increasing the fraction of Fe, the unit cell dimension progressively decreased, coupled with an increase of the plateaux pressure from 4 to 20 bar at 65 °C. All alloys displayed fast kinetics and no relevant changes on gravimetric capacity were reported, compared to the parent TiFe. In a recent study, Dematteis et al.[23] investigated the storage properties of a series of TiFe-based compounds and evaluated the effect of Mn and Ti substituting Fe in the alloy, by mapping Ti:Fe ratio from 1.0:1.0 to 1.0:0.9 and by varying the Mn content between 0 and 5 at.%. Samples were prepared at laboratory scale, similarly to Challet et al.[22]. Secondary minor phases of Ti$_4$Fe$_2$O-type, β-Ti-type or TiFe$_2$ were detected, always with abundance below 10 wt.%. The occurrence of the latter two compounds agrees with the TiFe phase diagram[24]. Ti-rich compositions, as compared to stoichiometric TiFe, form β-Ti-rich solid solution precipitates whereas, for Ti-poor compositions, the formation of TiFe$_2$ precipitates is expected. It is worth noting that the oxide was detected only in Ti-rich compositions, *i.e.* with Ti > 50 at.%, due to the high reactivity of Ti with oxygen, the latter being incorporated during sample preparation and/or already present in the starting materials[25]. Nevertheless, it was observed that a suitable quantity of secondary phases helped the activation process, but an excess causes a decrease in the gravimetric capacity. The addition of Mn in TiFe-based alloys increases the unit cell dimension, and a linear dependence was observed between the cell dimension and the equilibrium pressure for hydrogen sorption, as well as the pressure step between two plateaux in the pcT-curves and the hysteresis gap[21,22]. Authors concluded that the TiFe$_{0.85}$Mn$_{0.05}$ composition is the most promising for practical applications, thanks to an easy activation with 7 h of incubation time under H$_2$ at 25 bar and 25 °C, which results in fast absorption kinetics and a reversible capacity of 1.63 H$_2$ wt.%.

Thanks to the suitable H$_2$ sorption properties reported for TiFe$_{0.85}$Mn$_{0.05}$[22,23], this composition was selected as H$_2$-carrier for a plant designed to store up to 50 kilograms of H$_2$[26]. To operate the plant, the designated temperature is 55 °C. In fact, thanks to the observed equilibrium pressure at about 5.9 bar and 10.6 bar for the first and second absorption plateau, respectively[23], this alloy is suitable to absorb H$_2$ directly from a PEM-EL, with a supply pressure < 30 bar. Then, the H$_2$ could be released supplying a PEM-FC at 1 bar, owing to an equilibrium pressure of about 3.1 bar and 6.8 bar for the first and second desorption plateau, respectively[23]. The production and integration of a large-scale storage plant requires preliminary stages of prototyping, including a careful evaluation of the properties of the hydrogen storage material, which can differ moving from laboratory to industrial scale production. Indeed, the design can impose both technical and economic requirements that can drastically change the sorption properties of the selected materials. In particular, the purity of the starting raw materials will be dictated by the price and availability of parent elements. Processing techniques will differ, passing from a laboratory arc or induction melting to an industrial plant. The degree of control of the oxygen level at each step changes, due to a lower grade of vacuum during the synthesis and the occurrence of steps in air at high temperatures at an industrial level, compared to the controlled atmosphere maintained during the whole process at the laboratory scale. Finally, the economic design requirements limit the possibility after the synthesis to perform heat treatments at high temperatures and for long times, such as those described for laboratory scale processing in the literature[22,23]. This will lead to an industrial powdered material that, compared to that prepared at laboratory scale, is expected to be characterised by a higher oxygen content and a lower homogeneity, with the formation of gradients in composition or in secondary phases distribution, as well as by changes in microstructure and in surface properties.

Therefore, the first stage of prototyping the hydrogen storage plant was the characterization of a preliminarily batch of 5 kg of powder of a TiFe$_{0.85}$Mn$_{0.05}$ alloy produced under industrial conditions, to understand how the final hydrogen properties can be affected by the upscaling and to verify if the materials would be suitable for the final application. This work reports a detailed study of the phase composition, the microstructure and hydrogen sorption properties of an industrially produced TiFe$_{0.85}$Mn$_{0.05}$ alloy, comparing results with those obtained on a sample with the same overall composition prepared a laboratory scale, as reported in ref.[22,23]. The goal is to understand if the industrially prepared alloy is suitable to be used in the final plant, investigating its activation and operation at 55 °C, appropriate for the integration with a PEM-EL and PEM-FC, as well as its stability over cycling and its resistance to impurities in the gas supply.

## 2. Experimental

### 2.1. *Synthesis and processing*

An amount of 5 kg of TiFe$_{0.85}$Mn$_{0.05}$ was prepared from the parent elements by melting. Raw materials are electrolytic grade Fe and Mn and commercially pure (CP) Ti Grade 1 (minimal purity about 99.1 %). The chemical composition was obtained by taking a Ti-Fe-Mn master alloy with a nominal composition of 5.8 wt.% Mn and diluting the manganese to the target composition by additions of Fe and Ti. Dilution and melting were performed in a skull melter[27] under vacuum (10$^{-2}$ mbar). Then, the liquid alloy was poured in a water-cooled Cu-crucible. The resulting ingots were grinded in air in two steps. Firstly, the bulk was crushed in a jaw crusher to chunks, and then it was micronized in a hammer mill. The obtained powder was then poured in a sieve to select only the particle size with a dimension lower than 420 µm. Finally, the powder was delivered in air and then stored in a glovebox, under argon atmosphere with low concentration of oxygen and water (≤ 1 ppm).

## 2.2. Characterization

### 2.2.1 Chemical composition and oxygen content analysis

In the final powder form, the chemical composition of the alloy was determined with Microwave Plasma-Atomic Emission Spectrometry (MP-AES 4100 from Agilent), analysing 0.5 g, with a chemical pulping with 12 ml aqua regia and 2 ml HF. Oxygen and nitrogen contents were analysed with the elemental analyser Leco analyser ONH-863, while carbon and sulphur with the Leco CS-744.

### 2.2.2 Powder X-Ray Diffraction

For the as synthetized powder, Powder X-ray Diffraction (PXD) analysis was performed with a X'Pert Bragg-Brentano diffractometer, equipped with a Cu-Kα radiation and a X'celerator detector. The PXD patterns were acquired in air with an acquisition time of 1100 s per step, steps of 0.016°, from 35° to 120° in 2θ. After activation and cycling, the powders were analysed with a X'Pert Pro diffractometer in Debye-Scherrer geometry, equipped with Cu-Kα source. The powder samples were packed in glass capillaries with a diameter of 0.8 mm in a glovebox under Ar atmosphere. Measurements have been performed from 30° to 120° in 2θ, steps of 0.016° and a time per step of 400 s. Qualitative analyses were performed with the software X-Pert High Score, while Rietveld refinements of the crystal structures were carried out with the Maud software[28]. Pattern's background results to be noisy due to the fluorescence promoted by the presence of Fe in the sample and the use of Cu-Kα radiation.

### 2.2.3 Scanning Electron Microscopy

The study of the microstructure on the as synthetized sample was performed with the powder embedded in a conductive resin and polished, while morphological investigation was performed on loose powder. Analysis was carried out using a Field Emission Gun - Scanning Electron Microscopy (FEG-SEM) instrument Tescan 9000. The loose powder was also studied by performing Energy Dispersive X-ray Spectroscopy (EDX) measurements at different energy, *i.e.* 2, 5 and 15 keV, to perform Monte Carlo simulations with the software CASINO[29]. These were done to simulate the penetration depth of the beam inside the material. Finally, samples after cycling with H$_2$ as loose and embedded powders were studied with a SEM instrument Tescan Vega. On embedded samples, EDX elemental measurements were acquired at 20 keV.

### 2.2.4 Electron Probe Micro-Analysis

Metallographic examinations and elemental composition were obtained by Electron Probe Micro-Analysis (EPMA) with a Cameca SX100 instrument. Acceleration voltage and beam current were 15 kV and 40 mA, respectively. The as received powder was embedded in epoxy-resin and polished.

### 2.2.5 Volumetric Measurements by Sievert's Method

H$_2$ sorption properties were performed in different volumetric Sievert's apparatuses.

pcT-curves were registered with the instrument PCT-Pro by Setaram. About 2 g of sample were loaded in air and pure 6.0 hydrogen from Nippon gases was used. pcT-curves were registered maintaining the constant temperature with an electric furnace at 25, 40, 55 and 70 °C. Vacuum ($10^{-3}$ bar) conditions were applied at about 100 °C between a curve and another, to ensure a complete hydrogen desorption.

Two cycling tests were performed: a long-term cycling (250 cycles) and a poisoning experiment. In the first one, experiments were performed on a custom-made Sievert's differential pressure manometric apparatus, using 6.0 hydrogen from Linde. Approximately 200 mg of sample were loaded into a stainless-steel sample holder in a glovebox under Ar atmosphere. The temperature was regulated through an electric furnace. Measurements of hydrogen sorption as a function of time were performed at 55 °C and different pressures. Then, for the poisoning experiments, cycles were performed at 55 °C and between 25 and 1 bar, in another in-house built Sieverts-type apparatus. Some cycles were carried out using 6.0 purity hydrogen, referred to in this paper as *"pure H$_2$"*, from Nippon Gases. In order to check the effect of contaminants on sorption properties, fourteen additional cycles using 2.6 purity hydrogen, referred to as *"dirty H$_2$"*, from Nippon Gases. According to the supplier, the 2.6 purity hydrogen contained 50 ppm in volume of water, together with oxygen (< 100 ppmV) and nitrogen (< 3000 ppmV). About 1 g of powder was loaded into a stainless-steel sample holder in air.

### 2.2.6 Apparent density

Apparent density measurements were carried out inside a glovebox on as synthetized and activated powder samples. Tests were performed using a method like standard ISO 3953[30]. However, the small quantities of sample available (*e.g.* for the activated powder) did not allow complying with all the requirements of the standard procedure. A 25 cm$^3$ graduated cylinder was used, carefully cleaned and dried. Samples were poured inside the cylinder and then, the cylinder was manually put in vibration until a flat level surface was formed. The values of volumes were read on the cylinder, and those of the mass were determined by means of an analytical balance. Measurements were repeated three times and results were averaged.

## 3. Results and Discussion

### *3.1. Chemical and structural characterization*

Table 1 reports the chemical analysis for the as-synthetized powdered alloy.

Table 1: Chemical analysis on the as-synthetized alloy in powder form.

| *MP-AES* | | | | *Elemental analyzer* | | | |
|---|---|---|---|---|---|---|---|
| **Ti wt.%** | **Fe wt.%** | **Mn wt.%** | **Others wt.%** | **O wt.%** | **N wt.%** | **C wt.%** | **S wt.%** |
| 46.60 | 47.09 | 2.72 | < 3.28 | 0.233 | 0.006 | 0.015 | 0.007 |

Together with Ti, Fe and Mn, some impurities are present, linked to raw materials. In addition, a non-negligible amount of oxygen is observed. This could be associated to the purity of the starting Ti, which is the less pure element among the starting ones (section 2.1), and/or to material processing. Figure 1 shows the backscattered electron (BSE) image and the EDX-elemental maps for Ti, Fe, Mn and O acquired by FEG-SEM.

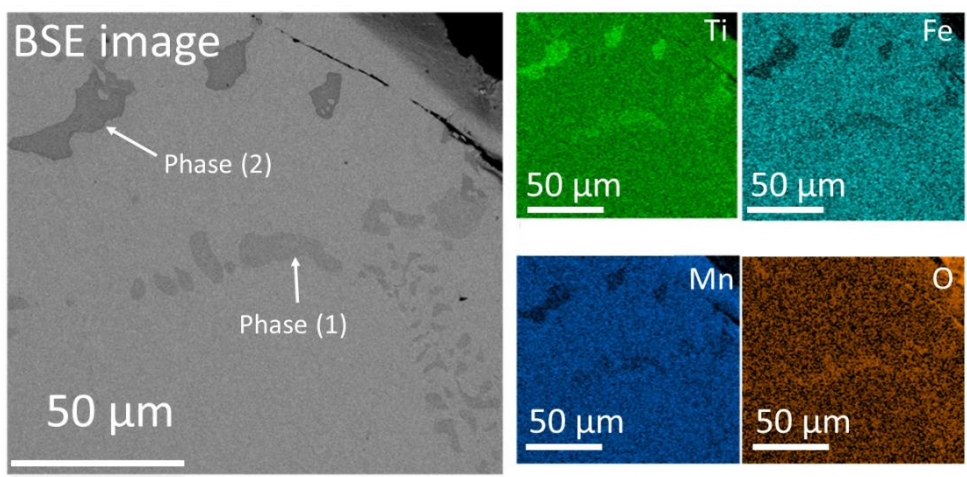

Figure 1: BSE image and EDX elemental maps for Ti, Fe, Mn and O, obtained by FEG-SEM on embedded as synthetized powder of the TiFe$_{0.85}$Mn$_{0.05}$ alloy prepared at industrial scale. The two zone darker than the matrix in the BSE image have been named Phase (1) and Phase (2) and correspond to Ti$_4$Fe$_2$O$_x$ and β-Ti$_{80}$(Fe,Mn)$_{20}$, respectively.

From the BSE image in Figure 1, it is possible to observe two zones darker than the matrix, named Phase (1) and Phase (2). From EDX maps, it appears that, compared to the matrix, both of them are richer in Ti and poorer in Fe and Mn, especially for the darkest one, *i.e.* Phase (2). In addition, it is also possible to observe from the EDX maps that Phase (1) presents a higher content of oxygen with respect to Phase (2). According to the EDX punctual analysis (Table 2), the composition of the matrix is confirmed to be TiFe$_{0.85}$Mn$_{0.05}$. Phase (1) presents an atomic ratio 2:1 in Ti and Fe, respectively, and it is linked to a Ti$_4$Fe$_2$O$_x$ phase. Phase (2) is the richest in Ti, and according to the elemental fractions reported in Table 2, it is linked to β-Ti$_{80}$(Fe,Mn)$_{20}$. For the Ti-rich sample, the occurrence of the β-Ti$_{80}$(Fe,Mn)$_{20}$ phase is expected, in agreement with the Ti-Fe phase diagram[24]. In contrast,

Ti$_4$Fe$_2$O$_x$ is not expected from the phase diagram and its formation is stabilized by oxygen impurities in raw materials, *e.g.* Ti, and/or from material processing[23,25]. Comparable results have been obtained performing the chemical investigation at microstructural level through EMPA analysis (Figure S1 and Table S1). As can be seen by the BSE image of Figure 1, Ti$_4$Fe$_2$O$_{0.4}$ (Phase (1)) is mainly present at grain boundaries, while the β-Ti$_{80}$(Fe,Mn)$_{20}$ (Phase (2)) is present as inclusions.

Table 2: Chemical compositions in at. % obtained by EDX analysis for Ti, Fe and Mn for the TiFe$_{0.85}$Mn$_{0.05}$ alloy prepared at industrial scale.

|  | Ti (at.%) | Fe (at.%) | Mn (at.%) |
|---|---|---|---|
| Matrix | 50.43 ± 3.96 | 45.78 ± 1.61 | 2.11 ± 0.44 |
| Phase (1) | 63.12 ± 1.15 | 34.00 ± 1.03 | 2.88 ± 0.12 |
| Phase (2) | 79.65 ± 0.04 | 16.97 ± 0.27 | 3.39 ± 0.32 |

The PXD pattern of the as-synthetized powder (Figure 2-a) was acquired to confirm the presence of the phases observed by SEM analysis.

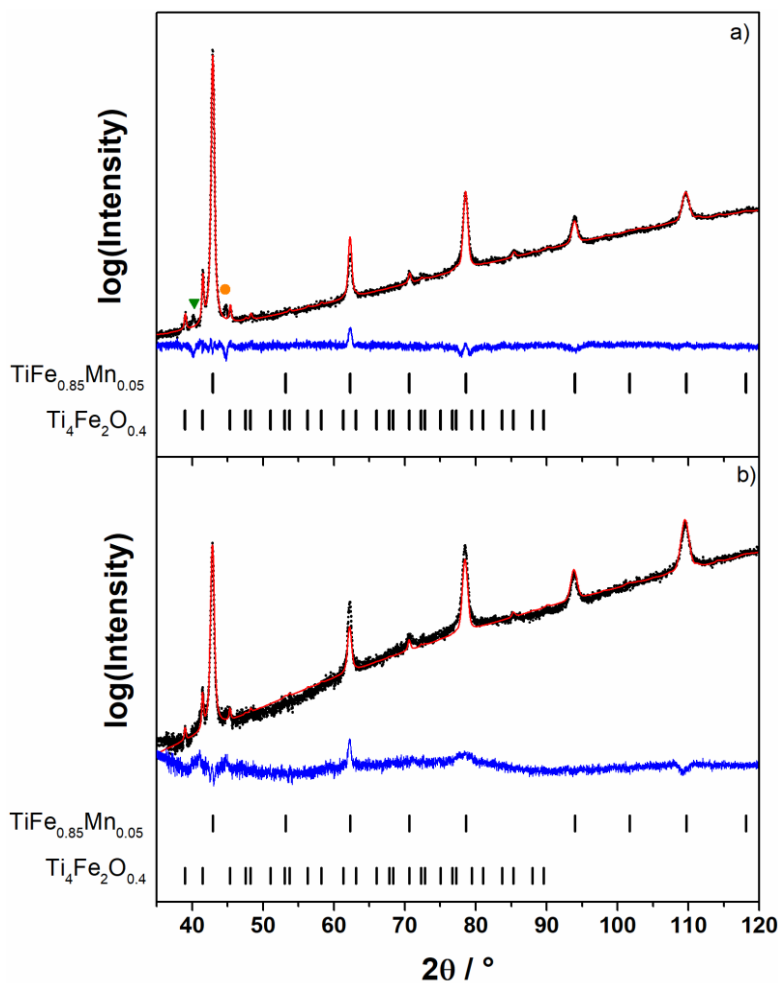

Figure 2: PXD patterns for the TiFe$_{0.85}$Mn$_{0.05}$ alloy prepared at industrial scale of (a) the as synthetized powder and (b) after activation. Dots represent experimental points, red lines the calculated patterns and blue lines the difference between them. Bars indicate diffraction peak positions for the single phases of TiFe$_{0.85}$Mn$_{0.05}$ and Ti$_4$Fe$_2$O$_{0.4}$, while the green

Barale et al. – TiFe$_{0.85}$Mn$_{0.05}$ alloy produced at industrial level for a hydrogen storage plant triangle refers to β-Ti$_{80}$(Fe,Mn)$_{20}$ and the orange circle to an unknown phase.

The phases detected by PXD (Figure 2-a) are in good agreement with the chemical analysis performed with the EDX. Indeed, diffraction peaks related to the presence of TiFe$_{0.85}$Mn$_{0.05}$, Ti$_4$Fe$_2$O$_x$ and β-Ti$_{80}$(Fe,Mn)$_{20}$ (green triangle) are observed. One diffraction peak, labelled with an orange circle and not assigned to the previous phases, was attributed to the presence of an unknown phase. Its stoichiometry is hard to assign, since only the main diffraction peak is visible with low intensity and no other phases were detected in the EDX analysis (Figure 1). The Rietveld refinement on the experimental patterns was performed to evaluate the lattice constants and phase abundance of Ti$_4$Fe$_2$O$_x$ and TiFe$_{0.85}$Mn$_{0.05}$. β-Ti$_{80}$(Fe,Mn)$_{20}$ is present in too small amount to be properly quantified. From the Rietveld refinement, 10 wt.% of Ti$_4$Fe$_2$O$_x$ is detected. The TiFe$_{0.85}$Mn$_{0.05}$ phase has a cubic CsCl-type structure, space group $Pm\bar{3}m$, with a cell parameter a = 2.984(1) Å. Both Ti and Mn partially substitute Fe, causing a lattice expansion compared to stoichiometric TiFe, a = 2.972 Å[5]. This result is in good agreement with literature for the same TiFe$_{0.85}$Mn$_{0.05}$ composition, *i.e.* 2.985(6) Å[22,23]. The oxide phase has a cubic structure, space group $Fd\bar{3}m$, with a cell parameter of 11.314(4) Å, in agreement with the literature[25]. The oxide Ti$_4$Fe$_2$O$_x$ has been reported in the literature with a variable composition (0.4 ≤ x ≤ 1.0) without significant changes in the lattice constant[25]. It is therefore not possible to determine accurately the stoichiometry of the oxide phase, from the cell parameter. The evaluation of the oxygen in the stoichiometry was instead performed by calculating the elemental amount of Ti, Fe, Mn, O, from the phase abundance obtained by the refinement for different oxygen stoichiometries of the oxide. For x = 0.4, the obtained amount of singe elements (*i.e.* Ti 50.11 wt.%, Fe 47.16 wt.%, Mn 2.52 wt.%, O 0.21 wt.%) is in good agreement with the results provided by the chemical analysis reported in Table 1. Thus, according to the results of the refinement and of the chemical characterization, the stoichiometry of the oxide phase was determined to be Ti$_4$Fe$_2$O$_{0.4}$.

Summarising, the industrial synthesis promotes the production of an inhomogeneous sample, with secondary phases present in higher amount compared to laboratory preparation[23]. Indeed, for the same composition, the industrially prepared material contains 10 wt.% of oxide phase Ti$_4$Fe$_2$O$_{0.4}$, compared to 2.4 wt.% of Ti$_4$Fe$_2$O$_x$ prepared at laboratory scale[23]. In addition, the latter has also a 2.8 wt.% of β-Ti$_{80}$(Fe,Mn)$_{20}$, resulting in a total amount of secondary phases equal to 5.2 wt.%[23], significatively lower than that observed in the powder prepared at industrial scale (i.e. 10 wt.% of the oxide phase).

### 3.2. *Hydrogen sorption properties*
### 3.2.1. *Activation*

The TiFe$_{0.85}$Mn$_{0.05}$ alloy prepared at laboratory scale can be easily activated after 7 hours of incubation time at 25 bar H$_2$ and 25 °C, displaying fast hydrogen absorption kinetics[23]. On the contrary, the industrially prepared material, maintained at the same conditions of temperature and pressure, did not absorb H$_2$ even after 3 days. This result confirms the impact of the synthesis method on the hydrogen sorption properties of MHs[6,31]. In this case, the material with the same nominal composition, prepared at either laboratory[23] or at industrial conditions, results in differences in phase fractions and microstructure, that strongly affect the hydrogen sorption properties, requiring the need to develop a specific activation procedure for each alloy. However, being the industrially produced powder the material to be used in a real storage system, the activation procedure needs to fit the plant constraints (*i.e.* maximum affordable pressure and temperature), to which the system is designed to operate in safe conditions. Moreover, it is also important to optimize the activation process from an economical point of view, limiting as much as possible the H$_2$ consumption to reduce the costs of the start-up of the plant. Therefore, for the industrially prepared material, a specific activation procedure was developed, trying to minimize the amount of hydrogen necessary and without exceeding 100 °C and 50 bar, parameters fixed by the storage plant design to ensure efficiency and safety.

The first attempt was done by performing loading and unloading of H$_2$ under isothermal conditions. In this attempt, the powder was maintained at 90 °C at 50 bar to promote absorption and at 1 bar to promote desorption. Unfortunately, after 30 cycles of 1 h each, the activation was not achieved, and Figure S2-a shows a schematic representation of one cycle by reporting pressure and temperature as a function of time. On the contrary, activation could be reached by heating the powder from 25°C up to 90 °C, keeping at 90 °C for 6 h under vacuum, followed by applying 50 bar of H$_2$ under isothermal conditions at 90 °C for 4 hours, cooling down in 2 hours to 25 °C and then keeping at 25 °C for 4 h. The total processing activation time is 16 h and requires a single hydrogen load (Figure S2-b). Analysing a small portion of the sample after the activation and vacuum pumping by PXD (Figure 2-b), it was observed that the oxide phase was still present in a percentage around 10 wt.%, as for the as synthetized powder (Figure 2-a). This result suggests that the activation process is linked to a change in the microstructure of the alloys, rather than to a change in the phase composition, as discussed later.

### 3.2.2. *Evaluation of the apparent density*

When studying a powder as H$_2$-carrier for a storage plant, it is of interest to evaluate the apparent density of the powder, which is a fundamental parameter for the design of the MH tank to calculate the amount of powder to allocate inside the reactor. Table 3 reports the values of apparent density obtained on the as synthetized and activated TiFe$_{0.85}$Mn$_{0.05}$ powder, prepared at industrial scale.

Table 3: Apparent density of the as synthetized and activated powder for the TiFe$_{0.85}$Mn$_{0.05}$ alloy prepared at industrial scale.

| Sample | Density g/cm$^3$ |
|---|---|
| As received | 3.94 |
| Activated | 3.28 |

The apparent density measured in this instance represents the bulk density of the tapped powder[30]. Typically, lower density values imply a hindered flowability of powders, due to irregular distribution of shapes/ratios and higher friction phenomena between particles. The size and its distribution also influence the density since finer powders often exhibit the abovementioned characteristics as well. In many cases, a decrease in density is evidence of the formation of smaller particles[30]. Thus, according to the values reported in Table 3, the lower value registered in the activated powder suggests the reduction of the particle size compared to the as synthesized one[32]. In fact, seen from the PXD patterns before and after the activation, shown in Figure 2, there are no evident variations in the phase fraction. The different powder morphology is linked to the processing in hydrogen that tends to reduce the particle size, due to the changes in volume occurring in the transformation from the hydride to the alloy[32]. Indeed, the role of the activation in hydrogenated powder is to crack the powder particles and promote the creation of new fresh surfaces[32]. It can be concluded that the activation procedure promotes the decrepitation of big particles into smaller ones, resulting in 17 % decrease of the final apparent density after activation. Finally, calculating the density from the structural information obtained by the Refinement and the chemical formula, the TiFe$_{0.85}$Mn$_{0.05}$ has a calculated crystal density of 6.45 g/cm$^3$. Thus, the apparent density of the powder before and after activation is about 60-50% of the crystal one.

### 3.2.3. *Thermodynamic study*

pcT-curves for the TiFe$_{0.85}$Mn$_{0.05}$ alloy prepared at industrial scale, obtained for hydrogen absorption and desorption, are reported in Figure 3.

Barale et al. – TiFe$_{0.85}$Mn$_{0.05}$ alloy produced at industrial level for a hydrogen storage plant

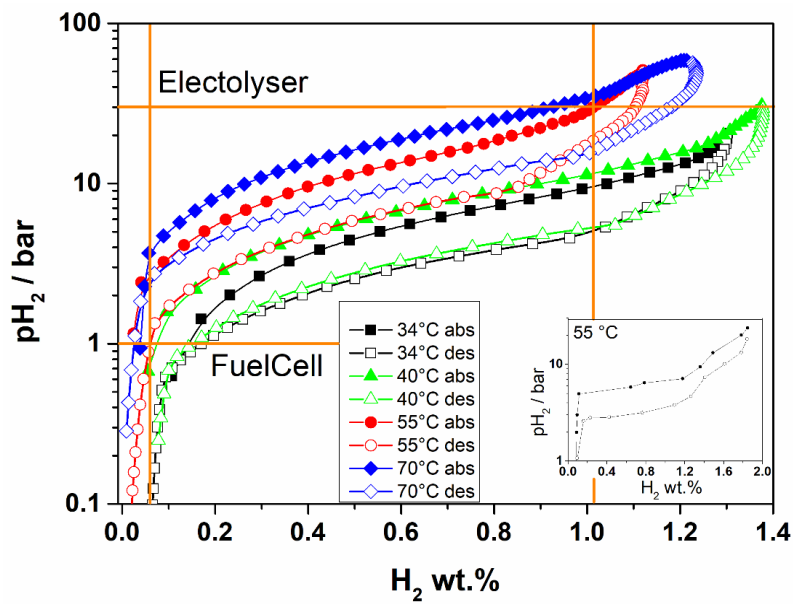

Figure 3: pcT-curves at 34-40-55-70 °C in absorption (full points) and desorption (empty points) for the TiFe$_{0.85}$Mn$_{0.05}$ alloy prepared at industrial scale, with solid lines as a guide for the eyes. The orange horizontal lines indicate the pressure of a PEM-EL with supply pressure at 30 bar and a possible desorption pressure at 1 bar for a PEM-FC. In the insert, the pcT curve at 55 °C, drawn from ref.[23], is reported.

Differently from what observed at a laboratory scale[22,23] (insert in Figure 3), all pcT-curves, in both absorption and desorption, display marked sloped plateaux. The two distinct plateau pressures are not observed. Due to the slope, the values of enthalpy (ΔH) and entropy (ΔS) for the hydrogen sorption reactions have been determined from the average of values obtained by applying the Van't Hoff equation at different H$_2$ wt.% (*i.e.* 0.4, 0.6 and 0.8, Table S3 and Figure S3). Results are reported in Table 4 as the average with the standard deviation, calculated by the Van't Hoff plots at the different H$_2$ wt.%, in comparison with data reported in ref.[23] for the same composition prepared at laboratory scale. Thermodynamic data are in the same range of values previously reported in the literature[23].

Table 4: Average enthalpy (ΔH) and entropy (ΔS) for the hydrogen sorption reactions, obtained in both absorption and desorption, for the TiFe$_{0.85}$Mn$_{0.05}$ alloy prepared at industrial scale. Values obtained for the same composition prepared at laboratory scale[23] are reported for comparison.

|  | Absorption | Desorption | Ref. [23] Absorption | Ref. [23] Desorption |
|---|---|---|---|---|
| ΔH kJmol$_{H2}^{-1}$ | -31.2 ± 0.8 | 31.2 ± 1.0 | 1$^{st}$ plateau: - 27.8<br>2$^{nd}$ plateau: -32.5 | 1$^{st}$ plateau: 30.6<br>2$^{nd}$ plateau: 35.2 |
| ΔS Jmol$_{H2}^{-1}$K$^{-1}$ | -115.7 ± 1.4 | 110.0 ± 1.3 | 1$^{st}$ plateau: - 99<br>2$^{nd}$ plateau: -121 | 1$^{st}$ plateau: 103<br>2$^{nd}$ plateau: 126 |

Considering the use of the material as hydrogen carrier to be supplied by a PEM-EL at 30 bar and to feed a PEM-FC at 1 bar (orange horizontal lines in Figure 3) at a working temperature of 55 °C, a reversible capacity of about 1.0 H$_2$ wt.% is expected (orange vertical lines in Figure 3), which is significantly lower than the value of 1.63 H$_2$ wt.% observed for the same composition prepared at the

Barale et al. – TiFe$_{0.85}$Mn$_{0.05}$ alloy produced at industrial level for a hydrogen storage plant laboratory scale[23], as evidenced in the insert of Figure 3. The marked slope and the reduced capacity for the industrially prepared material are linked to the absence of the annealing treatment after alloy synthesis and to the high fraction of secondary phases, which, as already discussed in section Chemical and structural characterization, is almost twice compared to that observed in the material prepared at the laboratory scale[23]. In fact, the annealing treatment induces a homogenization of the composition, promoting flat plateaux[21]. On the other hand, annealing is costly and difficult to perform at industrial level, especially for the laboratory process followed in ref.[22,23], *i.e.* one week annealing at 1000 °C. A progressive decrease of the H$_2$ storage capacity was also observed by increasing the quantity of secondary phases in ref.[23]. In this work, the discrepancy between the laboratory and industrial composition is due to the higher fraction of secondary phases observed. Indeed, both the oxide and β-Ti$_{80}$(Fe,Mn)$_{20}$ do not reversibly absorb hydrogen at the applied conditions[25], decreasing the overall reversible gravimetric capacity.

### 3.2.4. *Definition of plant conditions*

To define the conditions for the large-scale hydrogen storage plant, it is fundamental to determine the proper supply and release pressure, considering the system to be integrated with a PEM-EL (< 30 bar) upstream and a PEM-FC downstream (1 bar). So, the hydrogen sorption behaviour of the TiFe$_{0.85}$Mn$_{0.05}$ alloy prepared at industrial scale was investigated at different pressures, at the plant working temperature of 55 °C, evaluating five absorption pressures, *i.e.* 25, 20, 15, 10 and 5 bar and three desorption ones, i.e. 4, 2 and 1 bar (Figure 4).

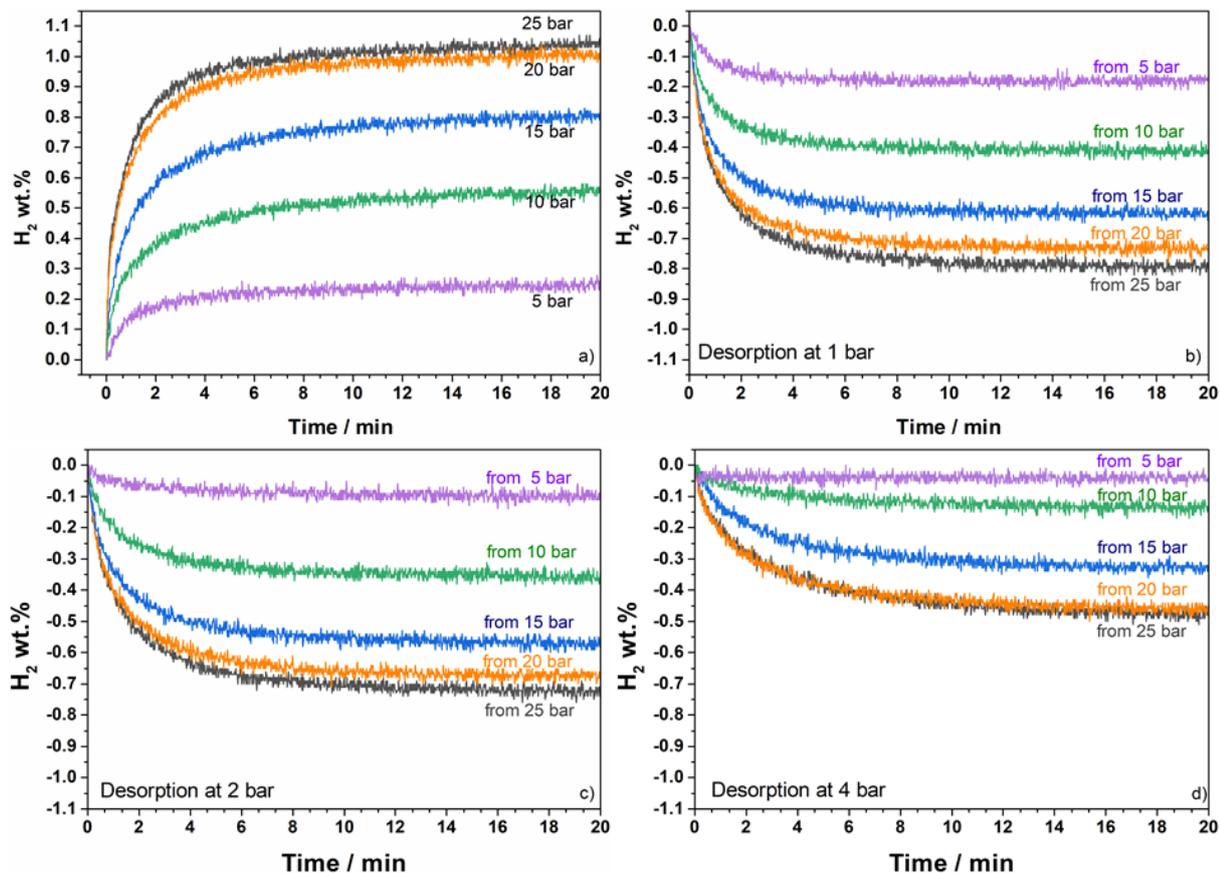

Figure 4: Absorption and desorption kinetic curves for the TiFe$_{0.85}$Mn$_{0.05}$ alloy prepared at industrial scale: (a) absorption at different pressures; (b) desorption from different pressures at 1 bar; (c) desorption at 2 bar; (d) desorption at 4 bar.

In general, for both reactions, the 90 % of the gravimetric capacity is processed in less than 10 minutes, highlighting a fast sorption rate. Considering the storage in absorption, in Figure S4, the total amount of hydrogen absorbed at different pressures at 55 °C is plotted as symbols over the measured pcT-curves. Values lays on the pcT-curve and the maximum absorption capacity increases with pressure, from 0.2 H$_2$ wt.% at 5 bar to 1.0 H$_2$ wt.% at 25 bar, in agreement with the thermodynamic study (Figure 3). For desorption, the capacity can be visualized thanks to the scheme presented in Figure S4. Arrows associate the absorbed quantities at loading pressure to the release of hydrogen in desorption at the equilibrium pressure 4, 2 and 1 bar, highlighted in the pcT-desorption curve. It can be noticed that, for selected conditions, the pressure levels exert almost no influence on the reaction driving force. Indeed, the last fraction of all the curves in Figure 4 is flat, suggesting that the reaction cannot proceed further due to thermodynamic constraints, rather than hindered kinetics. These results suggest that, for the plant at the designated working temperature of 55 °C, absorption can occur directly from a PEM-EL, between 25 and 30 bar, with a storage capacity of about 1.0 H$_2$ wt.%, in agreement with the thermodynamic study, and desorption can be achieved at 1 bar, directly supplying a PEM-FC, displaying a fast sorption rate.

### 3.2.5. *Cyclability study and resistance to gas impurities*

When studying an alloy for an industrial plant of this kind, the set of fundamental requirements includes its stability in performance over cycling and its resistance to possible hydrogen impurities. This makes necessary to check the expected lifetime of the material, in terms of reversible capacity and kinetic. performances. So, at the defined working conditions of temperature and pressure (*i.e.* 55 °C and 30 – 1 bar), several absorption and desorption cycles have been performed, to evaluate the stability of the hydrogen sorption properties of the TiFe$_{0.85}$Mn$_{0.05}$ alloy prepared at industrial level. As can be seen from Figure 5, the hydrogen storage capacity remains stable at about 1.0 H$_2$ wt.% over 250 cycles, observing even a slight increase (0.03 H$_2$ wt.%), and confirming the excellent stability of the alloy. Then, over the 250 cycles, kinetic performances are also unaffected, with single curves almost superimposable, as shown in Figure S5, reporting the hydrogen absorption curves taken every 50 cycles.

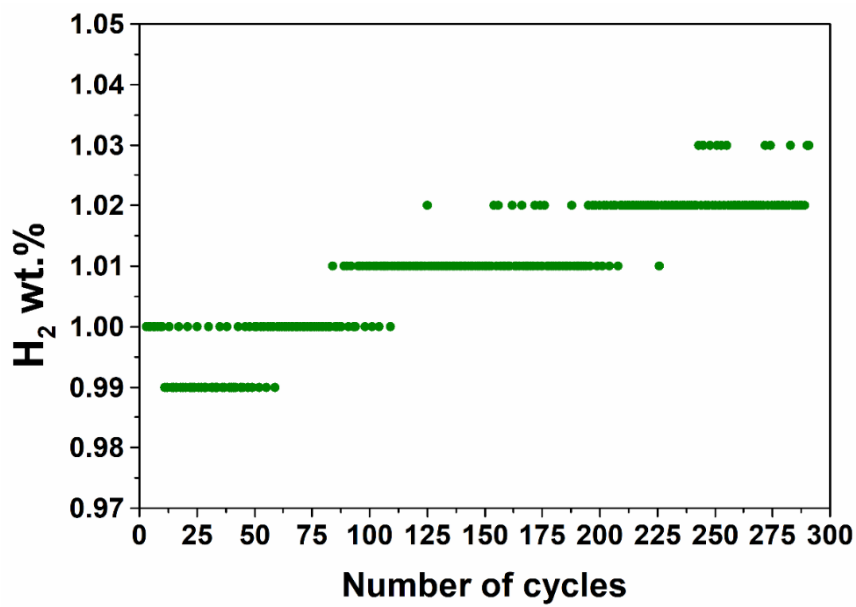

Figure 5: Gravimetric density (H$_2$ wt.%) as a function of the number of cycles for the TiFe$_{0.85}$Mn$_{0.05}$ alloy prepared at industrial scale. Analysis occurred at 55 °C absorbing at 25 bar and desorbing at 1 bar.

After more than 250 cycles with pure H$_2$, the powder was characterized by PXD and SEM. A comparison of the particle size of the powder after synthesis and hydrogen sorption cycles suggests that cycling promotes the reduction of the particle size. Indeed, by sieving after synthesis, the powder presents a dimension lower than 420 µm, while from a Secondary Electron SEM image on loose powder after 250 cycles (Figure S6), the average particle size results to be lower than 200 µm. The registered PXD pattern (Figure S7) shows that the main phase, TiFe$_{0.85}$Mn$_{0.05}$, is predominant, and the secondary phases detected on the as synthetised powder (Figure 2-a) are basically maintained, observing a reduction of the oxide and the evidence of a small new peak at about 35 ° in 2θ, linked

Barale et al. – TiFe$_{0.85}$Mn$_{0.05}$ alloy produced at industrial level for a hydrogen storage plant to an unidentified phase. It is worth noting that, performing an EDX-study on the matrix phase in an embedded powder after 250 cycles, the same TiFe$_{0.85}$Mn$_{0.5}$ composition has been detected (Table S8). Since, in the plant, H$_2$ is produced by an EL, impurities of water, oxygen and nitrogen are expected in the hydrogen flow. The effect of these contaminants on the hydrogen sorption properties have been therefore evaluated by performing absorption and desorption cycles using different grades of purity of H$_2$. The first ten cycles were carried using *"pure H$_2$"* (grade 6.0), followed by fourteen cycles performed with *"dirty H$_2$"* (grade 2.6). The amount of H$_2$ absorbed on cycling with pure and dirty H$_2$ is shown in Figure 6-a.

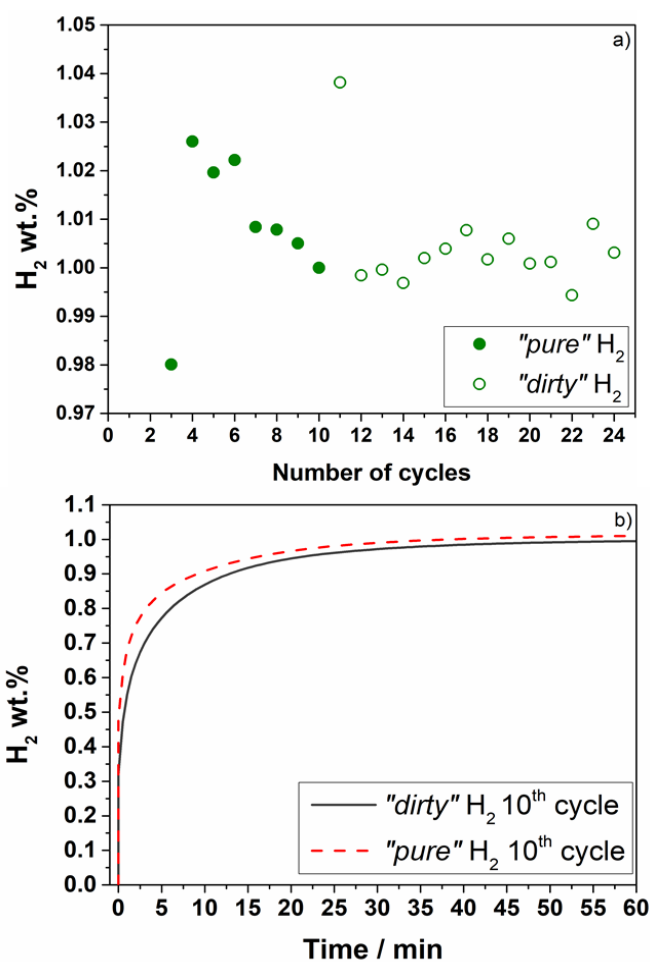

Figure 6: (a) Gravimetric density (H$_2$ wt.%) as a function of the number of cycles for the TiFe$_{0.85}$Mn$_{0.05}$ alloy prepared at industrial scale occurred at 55 °C absorbing at 25 bar and desorbing at 1 bar with *"pure"* and *"dirty"* H$_2$; (b) Gravimetric density (H$_2$ wt.%) as a function of time of the absorption curves acquired for the TiFe$_{0.85}$Mn$_{0.05}$ alloy prepared at industrial scale at the 10th cycle with *"pure"* and *"dirty"* H$_2$.

As shown in Figure 6-a, the experiment with the two grades of purity does not indicate any significant difference in the gravimetric capacity that remains stable at about 1.0 H$_2$ wt.%. On the other hand, the kinetic is slightly affected, as seen by comparing the absorption curve after 10 cycles with *"pure H$_2$"* and after 10 cycles (20 overall) with *"dirty H$_2$"* (Figure 6-b). Indeed, to reach 90 % of the total

H$_2$ capacity, 10 min are necessary with *"pure H$_2$"*, while 15 min are necessary with the *"dirty H$_2$"*, in which only the 80 % is processed in 10 min (Figure 6-b). Thus, it seems that impurities slow down the reaction rate, without affecting the maximum capacity upon cycling (Figure 6-a). Same considerations can be done for desorption (Figure S9).

The absorption reaction mechanism was studied through the Johnson–Mehl–Avrami (JMA) model[11], which allows to identify the rate determining mechanism of reaction by plotting the reacted fraction (α) as a function of time (t), according to Equation (1):

$$\ln(-\ln(1-\alpha)) = \ln k + m \ln t \quad \text{Equation (1)}$$

where *k* is a temperature-dependent rate parameter and *m* is a constant. Thanks to values tabulated by Hancock and Sharp[33], from the obtained value of *m*, it is possible to determine the rate-limiting mechanism. For the absorbing reactions with *"pure"* and *"dirty"* H$_2$, values of α were calculated and *ln(-ln(1 - α )* was plotted as a function of *lnt* (Figure S10), according to Equation (1). The evaluation of the m was made considering the entire range of α, resulting in values of 0.55 and 0.61 for *"pure"* and *"dirty"* H$_2$, respectively. Such values are quite similar and, according to ref.[33], can be linked to a diffusion mechanism as rate determine step. As it can be seen from Figure S10, all values are properly fitted in the whole range, suggesting that diffusion is the rate determining mechanism in the entire reaction. This implies that gas impurities do not affect the reaction mechanism, but it is reasonable to suppose that the slowing down is linked to the different partial pressure of H$_2$. In fact, as already reported in ref.[34], the decrease of the partial pressure of H$_2$, due to the presence and accumulation of impurities, promotes the slowdown of the reaction in comparison to the use of high purity H$_2$.

### 3.3. The role of oxygen in material processing and its influence on sorption properties

It is reported in the literature that the role of the oxygen introduced during the synthesis of TiFe-compounds, resulting in the formation of Ti-Fe-O ternary oxides, *e.g.* Ti$_4$Fe$_2$O$_x$ with 0.4≤x≤ 1, Ti$_3$Fe$_3$O or Ti$_{10}$Fe$_7$O$_3$, as secondary phases and/or as surfaces passive layer, is crucial in the sorption properties, starting already from their activation[12,25,35]. Indeed, ternary Ti-Fe-O oxides are always present in TiFe-based compounds, because of the oxygen immiscibility in the TiFe-based matrix and of the titanium high affinity with O[25]. Nevertheless, different results are reported in the literature regarding the ability of these oxides to be hydrogenated and, as a consequence, on their role in the H$_2$ sorption properties of the TiFe-compounds. Thus, in the following, the results obtained in these works are briefly summarized to help in the understanding the hydrogen sorption behaviour of the TiFe$_{0.85}$Mn$_{0.05}$ alloy prepared at industrial level.

Surface oxides as passive layer can be formed because of local concentration of oxygen at the surface, resulting in non-equilibrium Ti-Fe-O species[25,36]. Regarding the surface oxidation, it is reported in ref.[37,38] that the different electronic state of Fe in the matrix and in the surface oxide, when the material is heated, plays a catalytic role in the hydrogenation, promoting the dissociation of the hydrogen molecule. At this point the presence of grain boundaries between these oxides and passivated Ti(Fe,Mn) can act as fast channels for hydrogen diffusion, promoting alloy hydrogenation in bulk and by bulk expansion create cracks and fresh surfaces, helping in the hydrogenation[39–41]. The Ti$_3$Fe$_3$O and Ti$_{10}$Fe$_7$O$_3$ oxides were reported to be typical oxides present at the surface in TiFe-compounds; in particular, the Ti$_3$Fe$_3$O is reported as passive layer for the TiFe-alloys[42]. They do not absorb H$_2$, decreasing the storage capacity of TiFe-alloys[42–44]. On the other hand, they induce a reduction of the particle size, promoting hydrogenation. In fact, during hydrogen absorption and desorption cycles, the combination of a volume change of the matrix with the unchanged volume of the oxides enables particle size refinement[43]. Rupp et al.[25], by investigating the formation of ternary Ti$_{4-x}$Fe$_{2+x}$O$_y$ oxides in TiFe samples and their hydrogen sorption properties, concluded that these compounds are present in considerably high amount in commercial alloys, *i.e.* when industrial production is used, since high amount of oxygen can be introduced during processing. In particular, for Ti-rich compositions, the Ti$_4$Fe$_2$O$_{0.4}$ is formed. Ti$_{4-x}$Fe$_{2+x}$O$_y$ oxides were reported to absorb H$_2$ [25,35,45,46]. Their structure was investigated by neutron diffraction by Stioui et al.[35] and, more recently, by Density Functional Theory (DFT) calculation by Ha et al.[45]. A common result of investigations on Ti$_{4-x}$Fe$_{2+x}$O$_y$ oxides is that the oxygen content does not affect the cell parameter dimension, both in the hydrogenated and non-hydrogenated state, while it influences the amount of hydrogen absorbed. The higher the oxygen content, the lower the hydrogen storage capacity [25,35,45,46]. Several results were reported concerning the hydrogen sorption conditions (*i.e.* pressure and temperature) in Ti$_{4-x}$Fe$_{2+x}$O$_y$ oxides, but thermodynamic data are not available. Stioui et al.[35] reported easy hydrogenation of Ti$_4$Fe$_2$O at room temperature under few (not specified) bars of H$_2$ without need of activation. Also Rupp et al.[25] observed an easy hydrogenation between 20 °C and 100 °C under less than 1 bar of H$_2$ for the Ti$_4$Fe$_2$O$_{0.4}$, with a capacity of about 1.32 wt.% up to 5 bar, reaching even 1.76 wt.% in capacity hydrogenating the system at 5 bar, after cooling from 200 °C to room temperature. Hiebl et al.[46] reported a hydrogenation at 40 bar and 200 °C for Ti$_4$Fe$_2$O$_{0.46}$. It is important to point out that the easy activation and the hydrogenation at low pressures and temperatures observed in ref.[25,35] were obtained with the oxides in bulk form, while, in powder form, the hydrogenation is reported to be harder. Indeed, Rupp et al.[25] observed a deactivation of the oxide by powdering the bulk, in agreement with Matsumoto et al.[47] and observing hydrogenation only at 40 bar and 250 °C, in agreement with ref.[46]. This change in the sorption properties observed for the Ti$_4$Fe$_2$O$_{0.4}$ is not clear, but similar behaviour was observed for β-Ti$_{80}$Fe$_{20}$[25]. However, for the latter, hydrogenation in bulk form requires hard activation conditions, *i.e.* several cycles at 500 °C and 40 bar, while in powder form the hydrogenation occurs at 300 °C and 40 bar[25]. Finally, by preparing samples with different amount of Ti$_4$Fe$_2$O$_{0.4}$ and β-Ti$_{80}$Fe$_{20}$, together with the pure TiFe matrix, it was found that for high amount of Ti$_4$Fe$_2$O$_{0.4}$ at least 200 °C and 40 bar are required for the activation, while when β-Ti$_{80}$Fe$_{20}$ is present in higher amounts, hydrogenation occurs between 20 and 100 °C at 40 bar[25], in good agreement with ref.[47]. The hydrogenation of the secondary phases requires high pressures and temperatures, resulting in an overall reduction of the storage properties of the alloys, and when β-Ti$_{80}$Fe$_{20}$ is present in higher amount than the oxide, milder activation conditions are required[25]. Indeed, Nagai et al. in studying TiFeMn-samples, founds that the composition and the amount of the secondary phase, Ti-Fe-Mn-based, is crucial in the activation of the matrix. A proper fraction of secondary phases, even if does not absorb H$_2$, helps the activation thanks to its interface with the matrix acting as diffusion path for the hydrogen. The activation is empathised only when the secondary phases are even absorbing H$_2$ in the activation conditions[48]. Finally, the different thermal expansion between the matrix and the secondary phases helps the hydrogenation process, promoting the cracking of the alloys[19].

Taking into account the information from the literature, it is possible to describe the sorption behaviour of the TiFe$_{0.85}$Mn$_{0.05}$ sample prepared in this work at industrial level. In fact, the oxygen is affecting its production, with the formation of higher amounts of Ti$_4$Fe$_2$O$_{0.4}$ with respect to the same nominal composition prepared at a laboratory scale[22,23], in agreement with ref.[25]. The industrial preparation introduces a high amount of oxygen, because of less pure raw materials and material processing, that can promote the formation of oxides inclusions, together with the formation of a passive layer at the surface. The formation of a passive layer at the surface can be linked to the harsher activation observed if compared to ref.[23], considering that this passive layer is hindering the first hydrogenation, requiring a treatment in temperature and pressure to promote the diffusion through it[32]. Thus, the surface of the as synthetized material was studied by EDX analysis on loose powder with a FEG-SEM instrument, by changing the energy applied of the incident beam, *i.e.* 2, 5, and 15 keV. The rise of the energy of the incident beam implies a deeper penetration of the beam inside the material, allowing a progressive investigation of the composition of the material from the surface to the bulk[49]. The signal of the oxygen is taken as an indicator of the presence of the passive layer. Hence, if the passive layer is present, its value is expected to progressively decrease by increasing the energy, *i.e.* the penetration depth inside the matrix. The observed counts, normalized by the incident energy, are reported in Figure 7-a as a function of the emitted energy in keV, up to 1 keV. The EDX signal of oxygen occurs at about 0.5 keV and its decrease can be observed by increasing the beam energy applied, confirming a change in composition between the surface and the matrix.

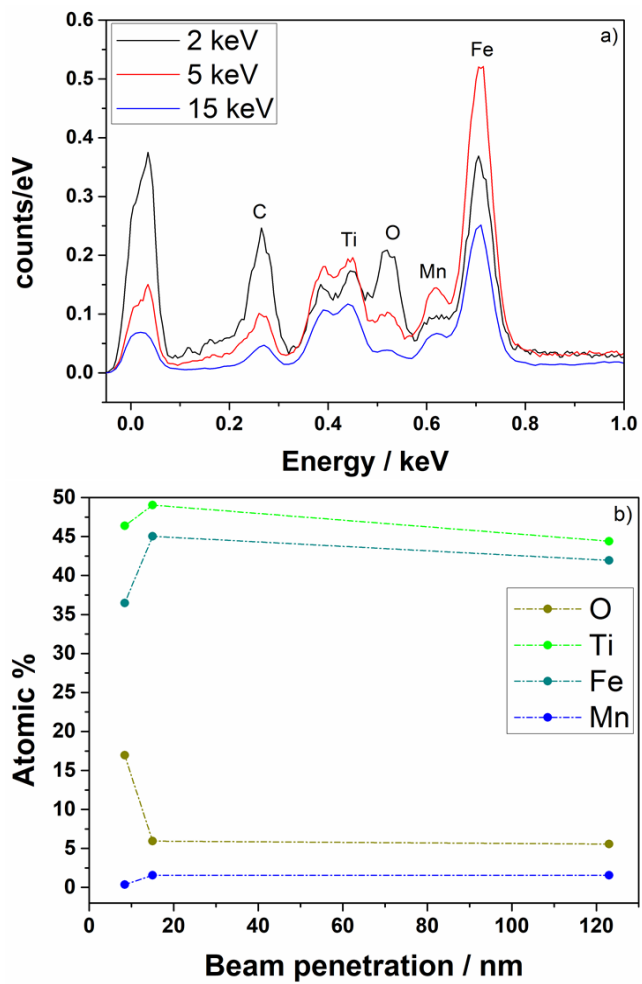

Figure 7: (a) EDX results, from FEG-SEM instrument, for the TiFe$_{0.85}$Mn$_{0.05}$ alloy prepared at industrial scale in terms of counts/eV as a function of the keV obtained with an experimental energy of 2, 5 and 15 keV; (b) the atomic percentage for Ti, Fe, Mn and O obtained by EDX analysis as a function of beam penetration at the 90 % of the emitted signal per each energy applied, *i.e.* 2, 5, and 15 keV.

The thickness of the passive layer was investigated by a Monte Carlo simulation with the software CASINO[29], estimating the penetration depth of the electronic beam as a function of the applied energy. An example of results of the electron penetration inside the material is displayed in Figure S11. As shown there, when an incident beam penetrates inside the material, the emitted signal comes from various depths. However, the more the beam penetrates inside the material, the lower the emitted signal is, passing from 90 % to 5 %. Thus, it was selected to consider the corresponding depth of the 90 % of the emitted energy, since it represents the zone, from which the majority of the information regarding the material comes from. Figure 7-b shows the elemental amount as a function of beam penetration at 90 % of the emitted energy, obtained with the Monte Carlo simulation. For the values reported in Figure 7-b, the highest amount of oxygen is linked to a beam penetration of 8.4 nm, linked to an applied beam energy of 2 keV, while at 5 keV and 15 keV, the depth is of 15 and 123 nm, respectively. The oxygen content decreases reaching a nearly stable value already at 15 nm, accompanied by a rise for Fe, Ti and Mn content. These results suggest the occurrence of a passive layer less than 15 nm thick, and the atomic percentage obtained from the EDX analysis (Figure 7-b) suggests a Ti:Fe:O atomic ratio equal to 3:2.5:1, i.e. similar to the Ti$_3$Fe$_3$O, already reported as passive layer of TiFe[42,44]. The presence of the passive layer cannot be confirmed through PXD analysis, since the penetration depth of the X-Rays calculated in the experimental conditions was estimated to be between 1 and 3 µm, significantly deeper than the value of 15 nm estimated from the Monte Carlo simulation for the EDX analysis. The presence of the Ti$_4$Fe$_2$O$_{0.4}$ at the surface of the powder cannot be excluded either, because of its high amount in the material. Indeed, since it was detected along grain boundaries (Figure 1), it can remain at the surface in some cases during the decrepitation of the powder.

On the basis of the literature data previously presented, regarding the sorption behaviour of Ti-Fe-O and β-Ti$_{80}$Fe$_{20}$ (β-Ti$_{80}$(Fe,Mn)$_{20}$ in this work), and the results observed for the TiFe$_{0.85}$Mn$_{0.05}$ prepared in this work at industrial level, Figure 8 shows a schematic illustrating the activation procedure (section 3.2.1).

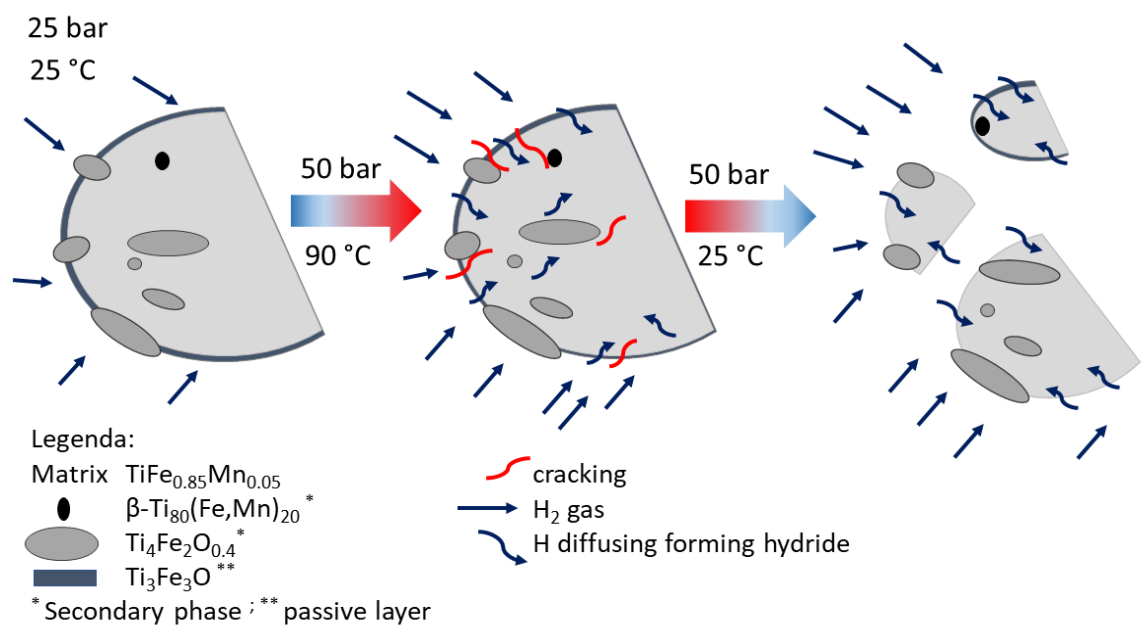

Figure 8: Schematic description of the activation mechanism for the TiFe$_{0.85}$Mn$_{0.05}$ alloy prepared at industrial scale.

At 25 °C and 25 bar (activation conditions of ref.[23]), the passive layer (Ti$_3$Fe$_3$O) hinders the first hydrogenation. Rising the temperature and pressure at 90 °C and 50 bar, respectively, promotes the diffusion of hydrogen through the passive layer, hydrogenating the matrix TiFe$_{0.85}$Mn$_{0.05}$. This process does not allow the hydrogenation of the secondary phases β-Ti$_{80}$(Fe,Mn)$_{20}$ and Ti$_4$Fe$_2$O$_{0.4}$, since higher temperature are required for a sample rich in oxide (*i.e.* 40 bar and 250 °C)[25]. However, both the passive layer, Ti$_3$Fe$_3$O, and the oxide, Ti$_4$Fe$_2$O$_{0.4}$, at the surface can play a crucial role during activation, by for instance, promoting a catalytic effect of the Fe[37,38] inducing the dissociation of H$_2$ and then, the hydrogenation of the matrix is facilitated due to the interface oxides-matrix[39–41]. The formation of new fresh surfaces occurs thanks to the variation in volume between the matrix and the oxide as unabsorbent species[43], promoting the cracking of the powder, enhanced also by the presence of β-Ti$_{80}$(Fe,Mn)$_{20}$[48]. It is worth noting that a more effective activation was obtained through a thermal cycle, suggesting a synergic effect between a variation in volume promoted by the hydrogenation and the different thermal expansion, especially of the Ti$_4$Fe$_2$O$_{0.4}$ presents in high amount and distributed along the grain boundary (Figure 1). During the activation, the particle size is reduced thanks to cracking promoted by the presence of the passive layer and secondary phases, causing the decrease in apparent density observed between the as synthesized and activated powder (Table 3).

The cracking of the powder allowed the formation of fresh surfaces to be hydrogenated, resulting in an activation procedure that still occurs in mild conditions and with only one thermal cycle, without exceeding the maximum affordable temperature and pressure of the plant.

The high amount of secondary phases and their inability to process hydrogen in the same condition of the TiFe$_{0.85}$Mn$_{0.05}$ industrially prepared[25], promoted a sensible reduction in the storage capacity, compared to the composition prepared at laboratory scale[23].

## 4. Conclusions

The TiFe$_{0.85}$Mn$_{0.05}$ alloy has been selected for a hydrogen storage plant and 5 kg in form of powder were prepared at industrial level by induction melting from the parent elements. It was found that the same composition prepared at laboratory[22,23] and industrial level, has a different microstructure, phase abundance and surfaces properties. These differences are caused by the synthesis itself and have a strong effect on the activation procedure, sorption capacities and kinetics. In detail, the activation is harder, as the powder needs to be heated and cooled in hydrogen atmosphere, and the storage capacity is significantly decreased. The thermodynamic is also affected, with a marked sloping plateau characterizing the pcT-curves, due to secondary phase abundance and sample inhomogeneity. Thermodynamic characteristic plays a key role in the actual H$_2$ sorption rate, as the driving force at the reaction conditions is lowered by the marked slope, hindering the performance. The storage capacity is of 1.0 H$_2$ wt.% at 55 °C and the sorption properties are maintained over 250 cycles. By using different grades of hydrogen purity, the storage capacity is not affected, while the accumulation of impurities, which affect the gas partial pressure, have a negative effect on the reaction rate. Finally, by linking all the structural and morphological characteristics, and phase abundance with the hydrogen sorption properties observed, the hydrogen sorption behaviour during the activation process was fully understood and explained. The properties determined in this work can be linked to the oxygen, introduced during the synthesis either from less pure Ti or from the material processing procedure. This promoted the formation of a significant fraction of oxides as secondary phases, Ti$_4$Fe$_2$O$_{0.4}$, and as passive layers, Ti$_3$Fe$_3$O. It can be concluded that metal hydrides are strongly influenced by the synthesis method and the industrial production results in a different hydrogen sorption behaviour that needs to be studied in order to understand if the material is still suitable for the final application. The mild activation method, the fast kinetic, and the resistance to gas impurities found for the TiFe$_{0.85}$Mn$_{0.05}$ alloy investigated in this work, confirm its suitability as H$_2$-carrier for a large-scale storage plant. Future work will be carried out in the realization of a prototype of the storage system, by producing about 50 kg of TiFe$_{0.85}$Mn$_{0.05}$ to investigate the alloy behaviour when integrated with a heat management.

## Acknowledgements

The project leading to this publication has received funding from the Fuel Cells and Hydrogen 2 Joint Undertaking (JU) under grant agreement No 826352, HyCARE project. The JU receives support from the European Union's Horizon 2020 research, Hydrogen Europe, Hydrogen Europe Research, innovation programme and Italy, France, Germany, Norway, which are all thankfully acknowledged. The authors wish to thank E. Leroy for EPMA analysis, F. Couturas for his help with hydrogenation experiments and G. Fiore for the performing of the FEG-SEM analysis.

This work honours the memory of Michel Latroche, a highly respected scientist and colleague, who substantially contributed to this work, but unfortunately passed away in December 2021.

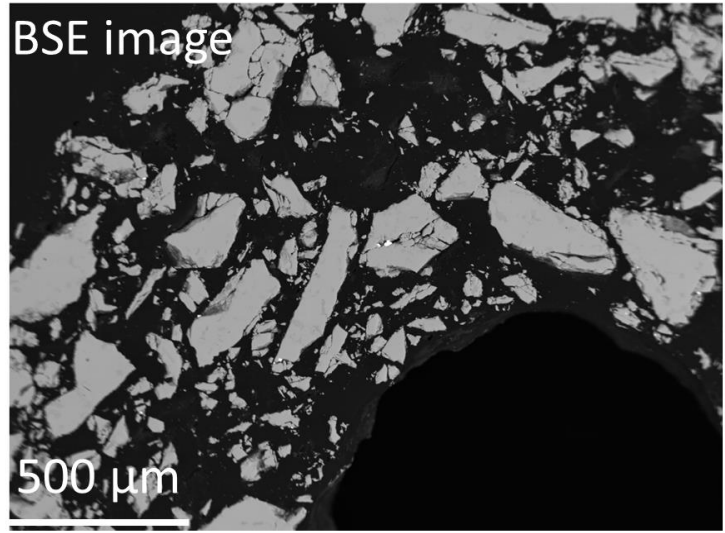

Figure S1 - BSE image of EMPA analysis on embedded as received powder.

|  | Ti (at.%) | Fe (at.%) | Mn (at.%) |
|---|---|---|---|
| Matrix | 52.35 ± 0.53 | 45.26 ± 0.96 | 2.40 ± 0.5 |
| Phase(1) | 65.08 ± 2.11 | 32.40 ± 2.02 | 2.52 ± 0.31 |
| Phase(2) | 81.71 ± 0.90 | 15.53 ± 0.95 | 2.76 ± 0.39 |

**Table S1 - Chemical composition obtained by EMPA analysis.**

Barale et al. – TiFe$_{0.85}$Mn$_{0.05}$ alloy produced at industrial level for a hydrogen storage plant

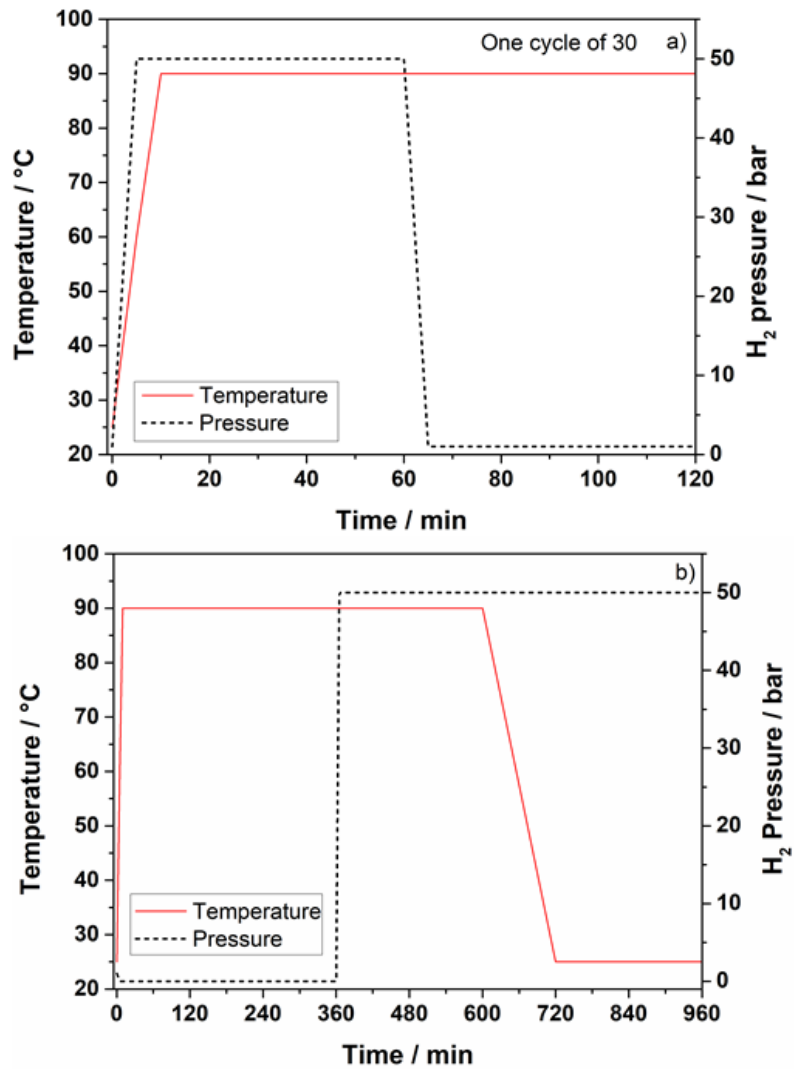

Figure S2 – Temperature and pressure trend as a function of time during the activation: (a) one cycle of the 30, performed in isothermal condition; (b) thermal cycle developed with one single loading of H$_2$.

Barale et al. – TiFe$_{0.85}$Mn$_{0.05}$ alloy produced at industrial level for a hydrogen storage plant

| H$_2$ wt.% considered | Absorption | | Desorption | |
| --- | --- | --- | --- | --- |
| | ΔH kJmol$_{H2}^{-1}$ | ΔS Jmol$_{H2}^{-1}$K$^{-1}$ | ΔH kJmol$_{H2}^{-1}$ | ΔS Jmol$_{H2}^{-1}$K$^{-1}$ |
| 0.4 | 32.1 | 115.9 | 32.0 | 109.7 |
| 0.6 | 30.7 | 114.2 | 31.6 | 111.4 |
| 0.8 | 30.7 | 116.9 | 30.1 | 109.0 |
| Average ± Std. Dev. | -31.2 ± 0.8 | -115.7 ± 1.4 | 31.2 ± 1.0 | 110.0 ± 1.3 |

Table S3 – Values of enthalpy and entropy in absorption and desorption obtained at 0.4, 0.6 and 0.8 H$_2$ wt.% from the pcT-curves.

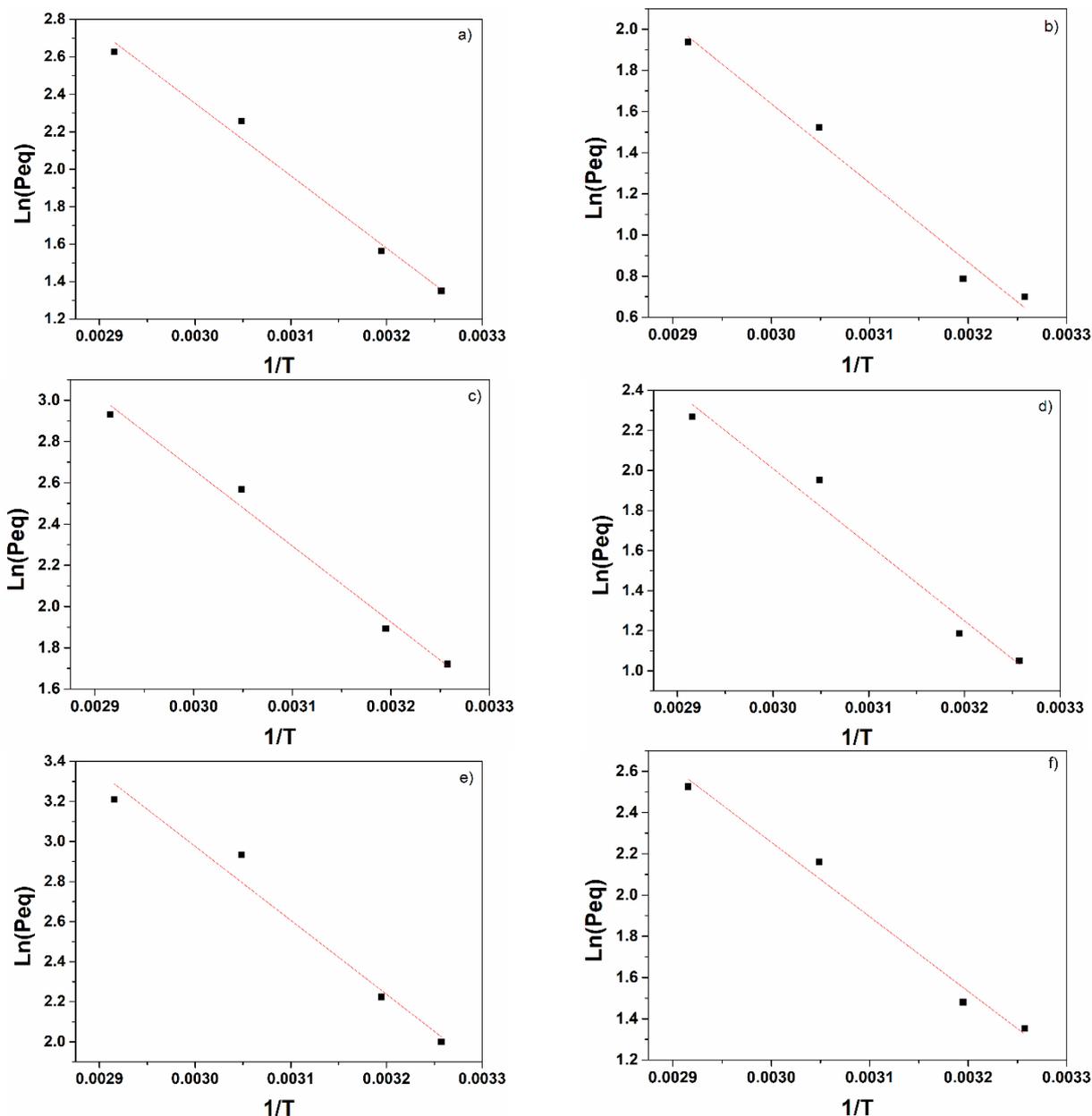

Figure S3 – Van't Hoff plots (a) absorption at 0.4 H$_2$ wt.%; (b) desorption at 0.4 H$_2$ wt.%; (c) absorption at 0.6 H$_2$ wt.%; (d) desorption at 0.6 H$_2$ wt.%; (e) absorption at 0.8 H$_2$ wt.%; (f) desorption at 0.8 H$_2$ wt.%.

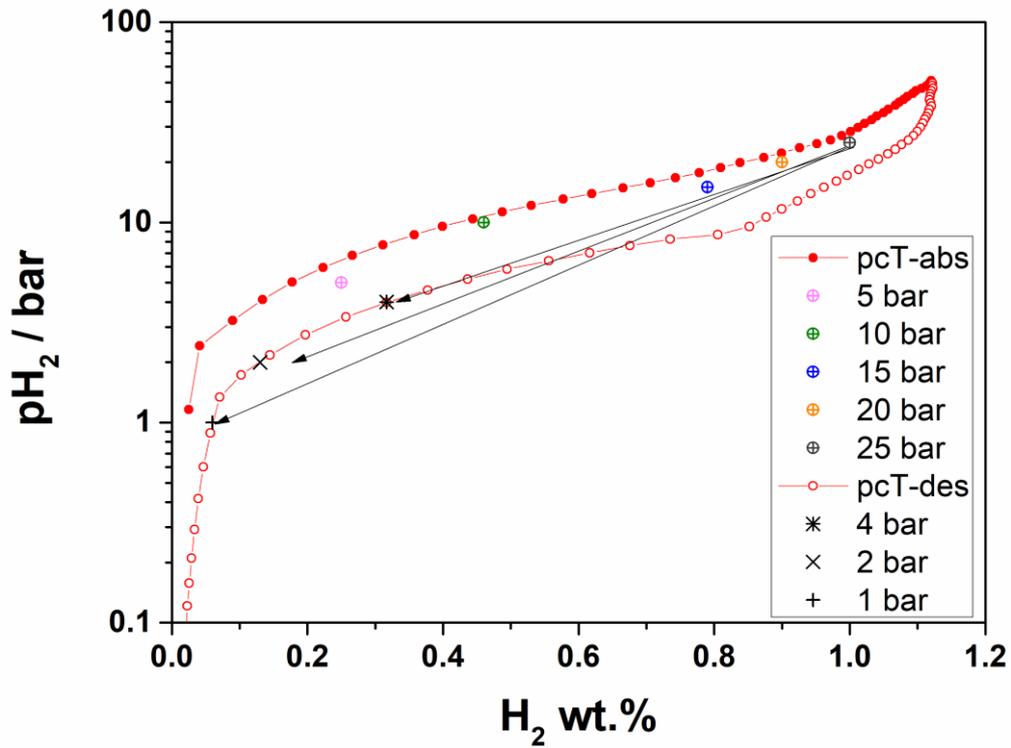

Figure S4 – pcT curves for absorption and desorption at 55 °C. Results obtained after absorption at different pressures (5, 10, 15, 20, 25 bar) and after desorption at different pressures (4, 2 and 1 bar) from a previous absorption at 25 bar.

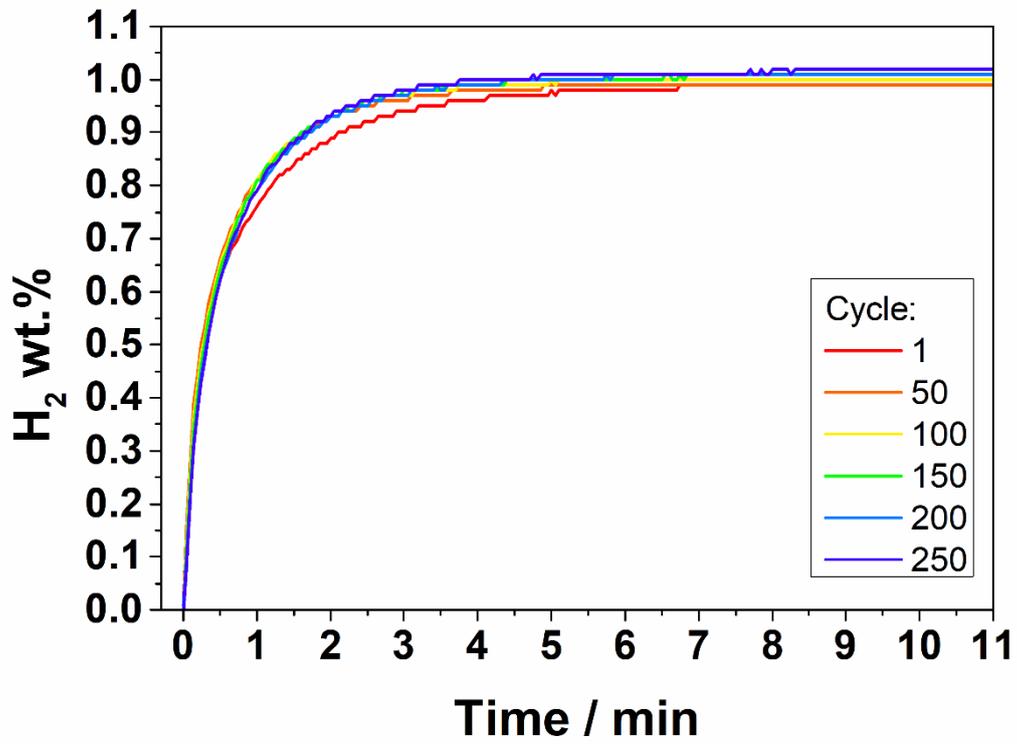

**Figure S5 - Gravimetric density (H$_2$ wt.%) as a function of time for hydrogen absorption registered in the cycling study taken every 50 cycles.**

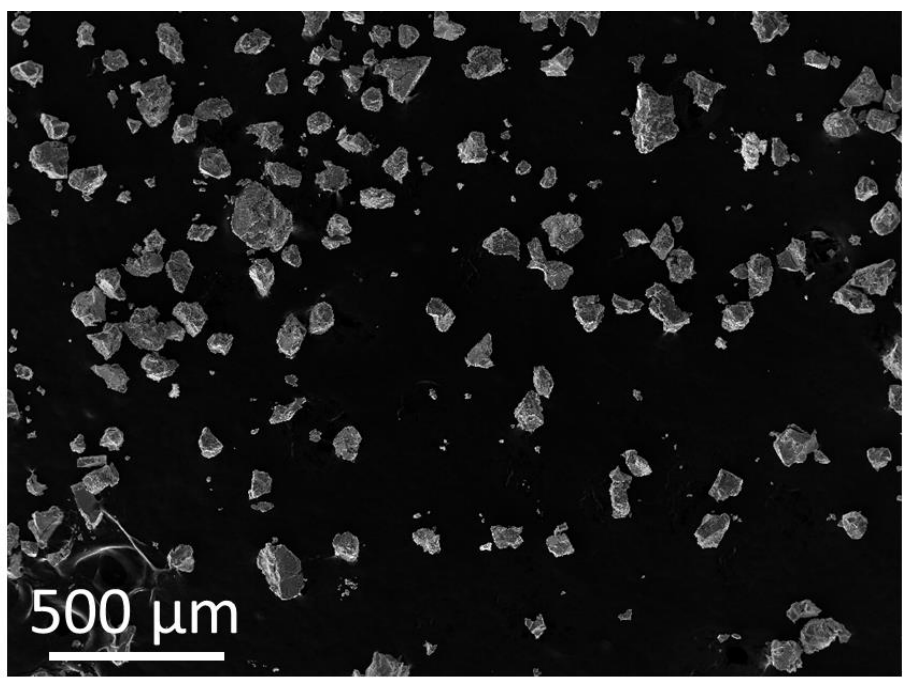

Figure S6 - SEM image in SE of the loose powder after 250 cycles under *pure* H$_2$.

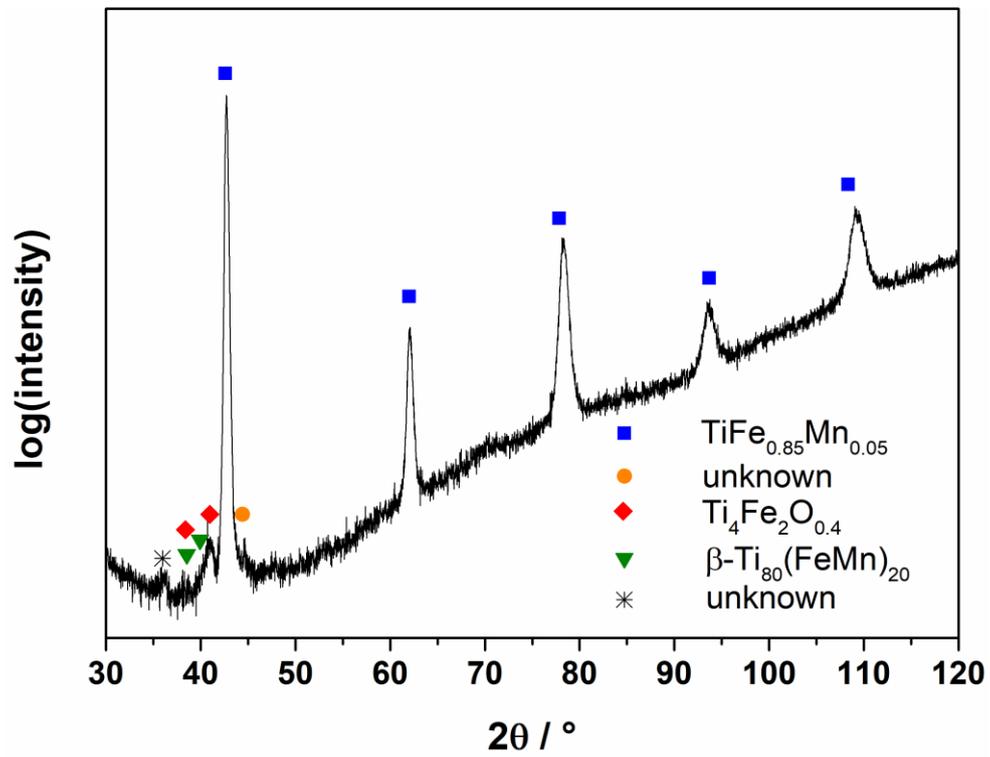

Figure S7 - PXD pattern of the powder after 250 cycles under *pure* H$_2$, with phases assignation.

|  | **Ti (at.%)** | **Fe (at.%)** | **Mn (at.%)** |
|---|---|---|---|
| Matrix | 52.06 ± 0.40 | 45.80 ± 0.96 | 2.14 ± 0.59 |

Table S8 - Chemical compositions in at.% obtained by EDX analysis for Ti, Fe and Mn on the powder after cycling.

Barale et al. – TiFe$_{0.85}$Mn$_{0.05}$ alloy produced at industrial level for a hydrogen storage plant

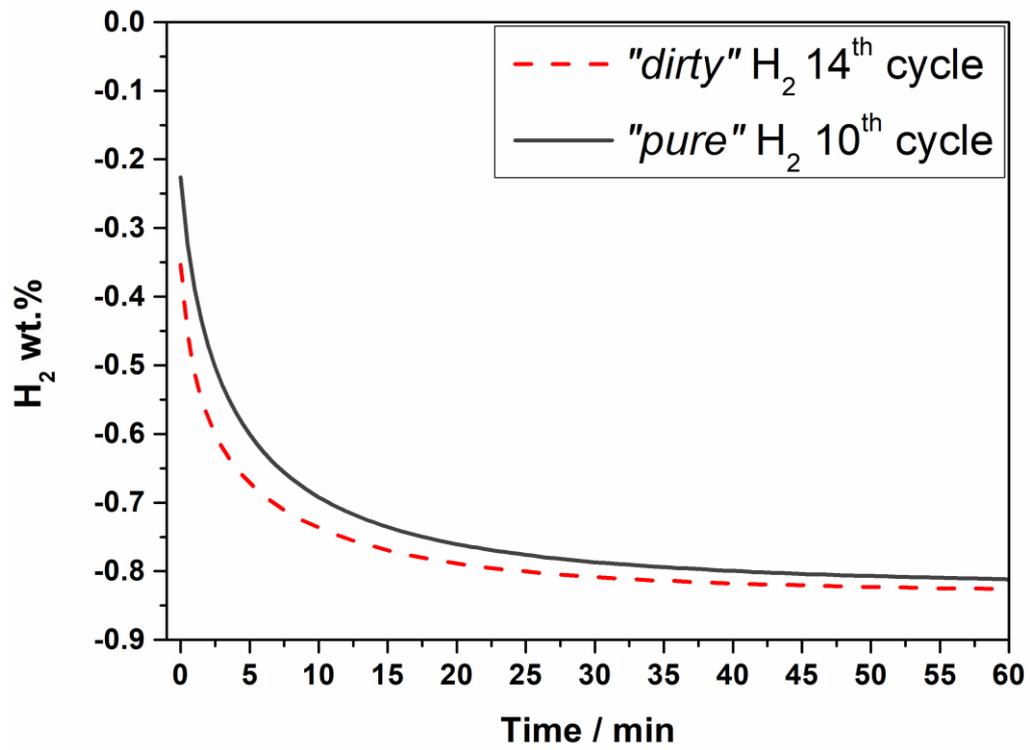

Figure S9 – The gravimetric capacity as a function of time in desorption for the 10$^{th}$ cycles with pure H$_2$ and the 14$^{th}$ one in dirty H$_2$.

Barale et al. – TiFe$_{0.85}$Mn$_{0.05}$ alloy produced at industrial level for a hydrogen storage plant

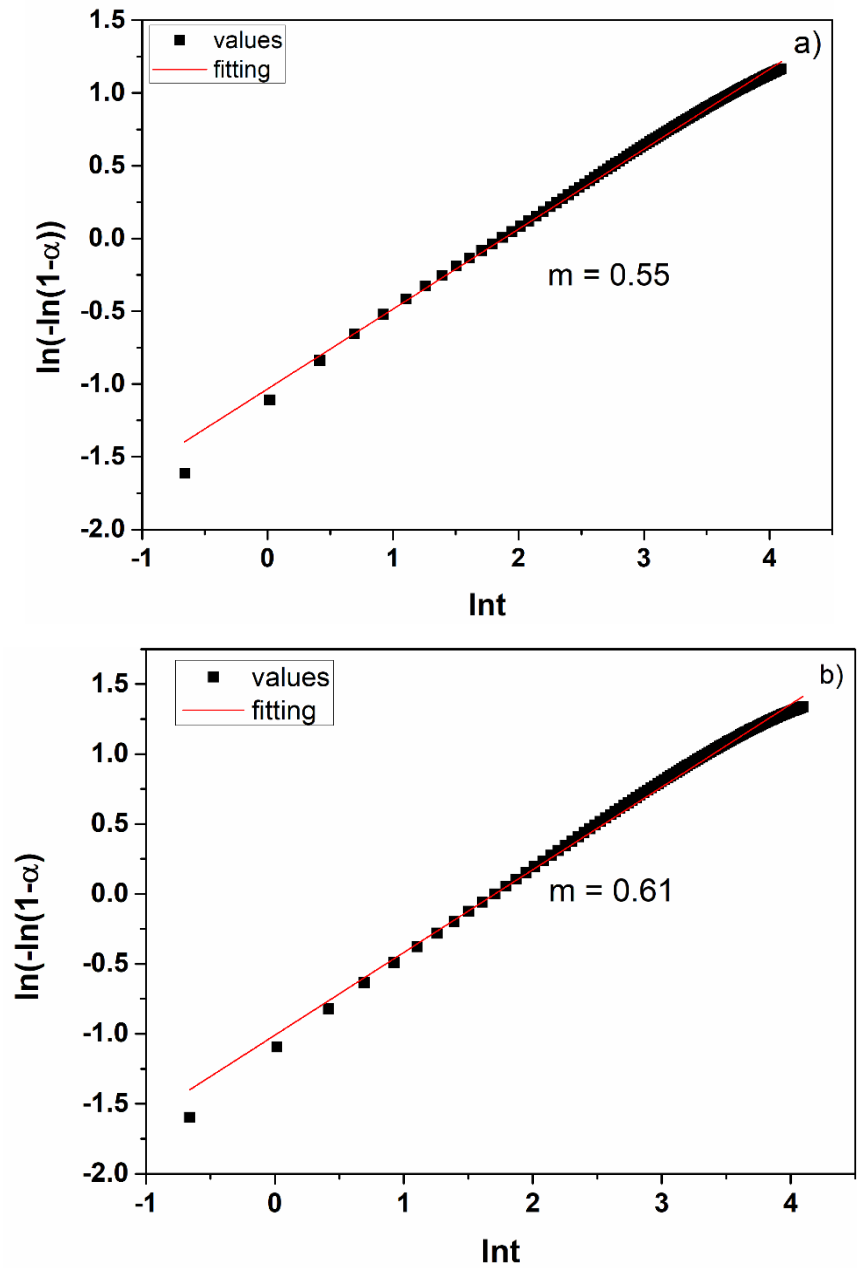

Figure S10 – Plots of ln(−ln(1 − α)) vs ln(t) for the absorption process (a) with pure and (b) with dirty hydrogen reporting the values of m obtained.

Barale et al. – TiFe$_{0.85}$Mn$_{0.05}$ alloy produced at industrial level for a hydrogen storage plant

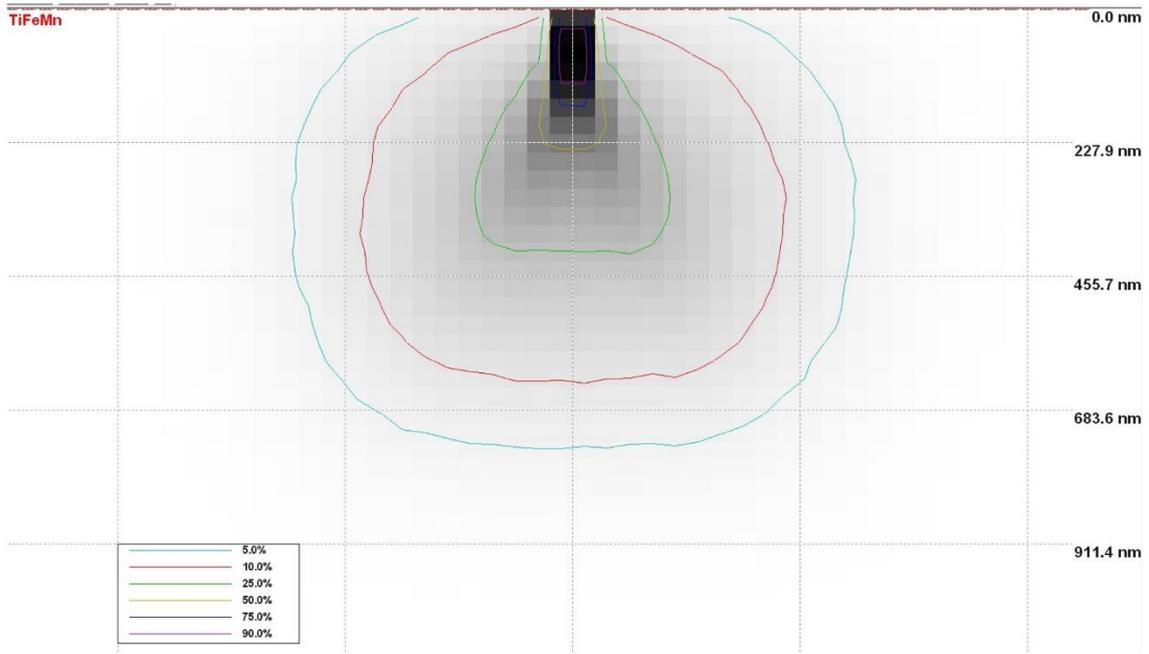

Figure S11 – Example of electron beam penetration inside the material obtained with the software CASINO for the measure at 15 keV, with the different color lines indicating the percentage of electrons inside the material. The penetration depth is reported in nm.